\documentclass[prl,,aps,superscriptaddress,twoside,twocolumn]{revtex4}

\usepackage{graphicx,epic,eepic,epsfig,amsmath,latexsym,amssymb,verbatim,revsymb,color}
\usepackage[noautoscale,vcentermath]{youngtab}
\usepackage{young}

\usepackage{theorem}
\newtheorem{definition}{Definition}

\newtheorem{lemma}[definition]{Lemma}

\newtheorem{corollary}[definition]{Corollary}

\def\squareforqed{\hbox{\rlap{$\sqcap$}$\sqcup$}}
\def\qed{\ifmmode\squareforqed\else{\unskip\nobreak\hfil
\penalty50\hskip1em\null\nobreak\hfil\squareforqed
\parfillskip=0pt\finalhyphendemerits=0\endgraf}\fi}
\def\endenv{\ifmmode\;\else{\unskip\nobreak\hfil
\penalty50\hskip1em\null\nobreak\hfil\;
\parfillskip=0pt\finalhyphendemerits=0\endgraf}\fi}
\newenvironment{proof}{\noindent \textbf{{Proof~} }}{\qed}

\newlength{\blank}
\mathchardef\ordinarycolon\mathcode`\:
\mathcode`\:=\string"8000
\def\vcentcolon{\mathrel{\mathop\ordinarycolon}}
\begingroup \catcode`\:=\active
  \lowercase{\endgroup
  \let :\vcentcolon
  }

\newcommand{\nc}{\newcommand}
\nc{\rnc}{\renewcommand}
\nc{\beq}{\begin{equation}}
\nc{\eeq}{{\end{equation}}}
\nc{\beqa}{\begin{eqnarray}}
\nc{\eeqa}{\end{eqnarray}}
\nc{\lbar}[1]{\overline{#1}}
\nc{\bra}[1]{\langle#1|}
\nc{\ket}[1]{|#1\rangle}
\nc{\ketbra}[2]{|#1\rangle\!\langle#2|}
\nc{\braket}[2]{\langle#1|#2\rangle}
\nc{\proj}[1]{| #1\rangle\!\langle #1 |}
\nc{\avg}[1]{\langle#1\rangle}
\nc{\Rank}{\operatorname{rank}\,}
\nc{\smfrac}[2]{\mbox{$\frac{#1}{#2}$}}
\nc{\tr}{\operatorname{Tr}}
\nc{\ox}{\otimes}
\nc{\dg}{\dagger}
\nc{\dn}{\downarrow}
\nc{\cA}{{\cal A}}
\nc{\cB}{{\cal B}}
\nc{\cC}{{\cal C}}
\nc{\cD}{{\cal D}}
\nc{\cE}{{\cal E}}
\nc{\cF}{{\cal F}}
\nc{\cG}{{\cal G}}
\nc{\cH}{{\cal H}}
\nc{\cI}{{\cal I}}
\nc{\cJ}{{\cal J}}
\nc{\cK}{{\cal K}}
\nc{\cL}{{\cal L}}
\nc{\cM}{{\cal M}}
\nc{\cN}{{\cal N}}
\nc{\cO}{{\cal O}}
\nc{\cP}{{\cal P}}
\nc{\cR}{{\cal R}}
\nc{\cS}{{\cal S}}
\nc{\cT}{{\cal T}}
\nc{\cX}{{\cal X}}
\nc{\cZ}{{\cal Z}}
\nc{\csupp}{{\operatorname{csupp}}}
\nc{\qsupp}{{\operatorname{qsupp}}}
\nc{\var}{\operatorname{var}}
\nc{\rar}{\rightarrow}
\nc{\lrar}{\longrightarrow}
\nc{\polylog}{\operatorname{polylog}}
\nc{\1}{{\openone}}

\nc{\RR}{{{\mathbb R}}}
\nc{\CC}{{{\mathbb C}}}
\nc{\FF}{{{\mathbb F}}}
\nc{\NN}{{{\mathbb N}}}
\nc{\ZZ}{{{\mathbb Z}}}
\nc{\PP}{{{\mathbb P}}}
\nc{\QQ}{{{\mathbb Q}}}
\nc{\UU}{{{\mathbb U}}}
\nc{\EE}{{{\mathbb E}}}
\nc{\id}{{\operatorname{id}}}

\nc{\be}{\begin{equation}}
\nc{\ee}{{\end{equation}}}
\nc{\bea}{\begin{eqnarray}}
\nc{\eea}{\end{eqnarray}}
\nc{\<}{\langle}
\rnc{\>}{\rangle}
\nc{\Hom}[2]{\mbox{Hom}(\CC^{#1},\CC^{#2})}
\nc{\rU}{\mbox{U}}

\nc{\ob}[1]{#1}

\nc{\LO}{\text{LO}}
\nc{\LOCC}{\text{LOCC}}
\nc{\cLOCC}{{\overline{\text{LOCC}}}}
\nc{\SEP}{\text{SEP}}
\nc{\PPT}{\text{PPT}}

\newcommand*{\anti}{{\,\yng(1,1)}}

\newcommand*{\complex}{\mathbb{C}}

\newcommand*{\half}{\frac{1}{2}}

\newcommand*{\col}{{\,\yng(1,1,1,1)}}
\newcommand*{\boxx}{{\,\yng(2,2)}}
\newcommand*{\hook}{{\,\yng(2,1,1)}}
\newcommand*{\sym}{\rm{Sym}}

\begin{document}

\title{Highly Entangled States With Almost No Secrecy}

\author{Matthias Christandl}
\affiliation{Faculty of Physics, Ludwig-Maximilians-Universit\"at M\"unchen, Theresienstr.~37, 80333 Munich, Germany}
\affiliation{Institute for Theoretical Physics, ETH Zurich, 8093 Zurich, Switzerland}

\author{Norbert Schuch}
\affiliation{Max-Planck-Institut f\"ur Quantenoptik, Hans-Kopfermann-Str.~1, D-85748 Garching, Germany}

\author{Andreas Winter}
\affiliation{Department of Mathematics, University of Bristol, Bristol BS8 1TW, U.K.}
\affiliation{Centre for Quantum Technologies, National University of Singapore, 2 Science Drive 3, Singapore 117542} %

\begin{abstract}
In this paper we illuminate the relation between entanglement and secrecy by providing the first example of a quantum state that is highly entangled, but from which, nevertheless, almost no secrecy can be extracted. More precisely, we provide two bounds on the bipartite entanglement of the totally antisymmetric state in dimension $d \times d$. First, we show that the amount of secrecy that can be extracted from the state is low, to be precise it is bounded by $O(1 / d)$. Second, we show that the state is highly entangled in the sense that we need a large amount of singlets to create the state: entanglement cost is larger than a constant, independent of $d$.
In order to obtain our results we use representation theory, linear programming and the entanglement measure known as squashed entanglement. Our findings also clarify the relation between the squashed entanglement and the relative entropy of entanglement.
\end{abstract}

\maketitle

\Yboxdim{4pt}

Entanglement is a quantum phenomenon governing the correlations between two parties.
It is both responsible for Einstein's ``spooky action at a distance'' as well as the security of quantum key distribution~\cite{BB84,E91}.
The universal resource for entanglement is the ebit, i.e.~the state $\ket{\psi}:=\frac{1}{\sqrt{2}} (\ket{00}+\ket{11})$~\cite{BBPS96}. Ebits are needed for teleportation, superdense coding and directly lead to secret bits. It is therefore natural to associate the usefulness of a quantum state with the \emph{distillable entanglement}, the amount of ebits that can be extracted from it asymptotically by local operations and classical communication (LOCC). The amount of ebits needed to create the state has been called \emph{entanglement cost}~\cite{BDSW96}, for which there is the formula~\cite{e-cost-paper}
\begin{equation} \label{eq:E_F-regularized} 
  E_C(\rho) = \lim_{n\rightarrow\infty} \frac{1}{n} E_F\bigl( \rho^{\ox n} \bigr), 
\end{equation}
with the \emph{entanglement of formation} $E_F(\rho)$~\cite{BDSW96}.

An important result relating to these quantities has been the discovery of bound entanglement, that is of states that need ebits for their creation but from which no ebits can be extracted: $E_C(\rho)>0$ and $E_D(\rho)=0$~\cite{bound-e-paper}. A recent surprise has been the realization that there exist bound entangled states from which secrecy can be extracted~\cite{horodecki-2005-94}, a result that overthrew previous beliefs that secrecy extraction and entanglement distillation would go hand in hand.
The amount of key that can be extracted from a quantum state is known as the \emph{distillable key} $K_D(\rho)$,
and a fundamental question at this point is the following.
Are there states requiring key to create them but from which no secret key can be distilled? Even the weaker form, whether there exist states with
$E_C(\rho) > 0$ but $K_D(\rho) = 0$, seems too difficult at the moment. Here we show that in an asymptotic
sense the answer is yes. In the spirit of~\cite{RennerWolf:boundkey}, we show that there exists a family
of states with constant lower bound on their entanglement cost, but arbitrarily small distillable key. 

Our example is the well-known antisymmetric state $\alpha_d$ in $\CC^d \ox \CC^d$, that is the state proportional to the projector onto the antisymmetric subspace. Our main results are:
\begin{align}
  \label{eq:EC-lower}
  E_C(\alpha_d) &\geq \log_2 \frac{4}{3}, \text{ and} \\
  \label{eq:KD-upper}
  K_D(\alpha_d) &\leq \left. \begin{cases}
                               \phantom{\half}\log_2\frac{d+2}{d} & \text{ if } d \text{ is even}    \\
                               \half\log_2\frac{d+3}{d-1} & \text{ if } d \text{ is odd}
                             \end{cases}
                       \right\} = O\left(\frac{1}{d}\right).
\end{align}

Being an extreme point of the set of Werner states, some entanglement measures 
have been computed for $\alpha_d$ previously~\cite{Rains2001, audenaert-2001-87}, although 
entanglement cost has defied its calculation. The only exception was Yura's tour de 
force calculation in which he proved that $E_C(\alpha_3)=1$~\cite{Yura:E_C}. 
Perhaps researchers had also lost interested in the problem since the additivity 
conjecture of entanglement of formation~\cite{Shor:equivalences} would have implied 
that $E_C(\alpha_d)=1$, as it is easy to see that for all $d$, 
$E_F(\alpha_d) = 1$. Now this conjecture is known to be
false~\cite{Hastings}, and we thus believe that our result also sheds light on 
the old problem of calculating the entanglement cost and the cases in which at 
least some weak form of additivity might hold. 
We emphasize that the value of $\log_2\frac{4}{3}$ is only a lower bound, and that 
it is quite conceivable that $E_C(\alpha_d) = 1$.

For each of the two bounds, we introduce a new technique that may be of interest in its own right. 
We start by deriving the upper bound on the distillable key. The argument consists of two steps. First we show that squashed entanglement is an upper bound on the distillable key. Then we provide an upper bound on squashed entanglement for the antisymmetric states. 
To lower bound the entanglement cost, we relax the calculation of $E_F(\alpha_d^{\otimes n})$ first to a semidefinite programme, and then reduce this with the help of representation theory -- for the first time using the concept of plethysm in quantum information theory -- to a linear programme. Finally, we find a feasible point of the dual programme, resulting in our lower bound. Along the way, we recover Yura's result due to a representation-theoretic simplification in $d=3$.

\medskip\noindent
{\bf Upper bound on distillable key.}
\label{sec:key}
The formal definition of the distillable key is
$$K_D(\rho_{AB}) = \lim_{\epsilon \rightarrow 0} \lim_{n \rightarrow \infty} \sup \left\{ \frac{m}{n}: \| \Lambda_n(\rho^{\otimes n})-\gamma_m \|_1 \leq \epsilon \right\},$$
where the maximisation extends over LOCC protocols $\Lambda_n$ and states $\gamma_m$ that contain $m$ bits of pure secrecy. More precisely, $\gamma_m=U \sigma_{AA'BB'}U^\dagger$ for some controlled-unitary
$U=\sum_{i=1}^{2^m} \proj{ii} \otimes U_i$ and
$\sigma_{AA'BB'}=\proj{\Phi}_{AB} \otimes \sigma_{A'B'}$, where
$\ket{\Phi}=\frac{1}{\sqrt{2^m}} \sum_{i=1}^{2^m} \ket{i}\ket{i}$ is
the maximally entangled state of rank $2^m$~\cite{horodecki-2005-94}. 
Recall the squashed entanglement~\cite{squashed}
$$E_{\rm sq}(\rho_{AB})=\inf_{\rho_{ABE}: \rho_{AB}=\tr_E \rho_{ABE}} \half I(A;B|E)_{\rho},$$
where $I(A;B|E)=H(AE)+H(BE)-H(ABE)-H(E)$ is the quantum conditional mutual information with $H(\cdot)$ the von Neumann entropy. The following result was first announced in~\cite{christandlPhD}.
\begin{lemma} \label{lemma:squashedupper} 
  For all $\rho_{AB}$, $K_D(\rho_{AB}) \leq E_{\rm sq}(\rho_{AB})$. 
  \end{lemma}
\begin{proof}
Let $\Lambda_n$ be an LOCC protocol such that 
$  \|\Lambda_n(\rho^{\otimes n})- \gamma_m \|_1 \leq \epsilon.$
Since squashed entanglement is
a monotone under LOCC~\cite{squashed} and asymptotically continuous~\cite{AliFan04},
$E_{\rm sq}(\rho^{\otimes n}) \geq E_{\rm sq}\bigl(\Lambda(\rho^{\otimes n})\bigr) 
                          \geq E_{\rm sq}(\gamma_m)-\delta(\epsilon) n$,
where $\delta(\epsilon)$ approaches zero as $\epsilon$ decreases~\footnote{Note that
the dimensions of $\gamma_m$ can be assumed to be exponential 
in $n$, which is important in the bound~\cite{AliFan04}.}.
Recall the form of the state $\gamma = \gamma_m$: In order to show
that $E_{\rm sq}(\gamma_m) \geq m$, consider an extension $\sigma_{AA'BB'E}$
of $\sigma_{AA'BB'}$ and the induced extension 
$\gamma_{AA'BB'E}=(U \otimes \openone_E) \sigma_{AA'BB'E} (U^\dagger \otimes \openone_E)$ of $\gamma_{AA'BB'}$.
Clearly, $H(AA'BB'E)_\gamma=H(AA'BB'E)_\sigma= H(A'B'E)_\sigma=H(A'B'E)_{\sigma_i},$
with $\sigma_{A'B'E, i}:= (U_i \otimes \openone_E) \sigma_{A'B'E} (U_i^\dagger \otimes \openone_E)$. 
Furthermore, $H(E)_{\sigma_i}=H(E)_\sigma \quad \textrm{and} \quad H(AA'E)_\gamma=H(A)_\gamma+\sum_i p_i H(A'E)_{\sigma_i}$ and similarly for $H(BB'E)_\gamma$. Altogether this gives
$I(AA';BB'|E)_\gamma \geq H(A)_\gamma+H(B)_\gamma+\sum_i p_i I(A';B'|E)_{\sigma_i} \geq 2m$,
where we used the non-negativity of the quantum mutual information. This shows that $E_{\rm sq}(\gamma_m)\geq m$
and therefore $E_{\rm sq}(\rho) \geq \frac{m}{n}-\delta(\epsilon)$. Choosing a sequence of protocols $\Lambda_n$ that achieves the distillable key, the r.h.s.~converges to $K_D(\rho)$.
\end{proof}

\medskip
In order to find an upper bound on distillable key, it thus suffices to upper bound squashed entanglement:
\begin{equation} \label{eq:squashed} 
  E_{\rm sq}(\alpha_d) \leq \begin{cases} \phantom{\half} \log_2 \frac{d+2}{d} \quad  & d \text{ even }\\
                                       \half          \log_2 \frac{d+3}{d-1}  &  d \text{ odd. }
                        \end{cases}
\end{equation}
Let $P_k$ be the projector onto $\wedge^k(\complex^d)$, the antisymmetric subspace of $k$ systems with local dimension $d$. Note that $d_k:=\dim \wedge^k(\complex^d)= \binom{d}{k}$ and $\alpha_d = \frac{P_2}{d_2}$.
We make use of the fact that $\alpha_{ABE}:=\frac{P_k}{d_k}$ is an extension of 
$\alpha_{AB}$ if $E$ consists of $k-2$ particles. In that case,
$I(A;B|E)_{\alpha}=H(AE)_\alpha+H(BE)_\alpha-H(E)_\alpha-H(ABE)_\alpha
=\log_2 \frac{d_{k-1}^2}{d_{k-2}d_k}=\log_2 \frac{k}{k-1}\frac{d-k+2}{d-k+1}$.
The values in \eqref{eq:squashed} are then obtained by minimising over $k$; the minima are reached for $\frac{d}{2}+1$ and $\frac{d+1}{2}$, for even and odd $d$, respectively.

\medskip
This is remarkable as the regularised relative entropy of entanglement with respect to PPT states, the Rains bound, and the logarithmic negativity are all equal to $\log_2 \frac{d+2}{d}$ for $\alpha_d$~\cite{Rains2001, audenaert-2001-87}. In the light of these results we are tempted to conjecture that $E_{\rm sq}(\alpha_d) = \log_2 \frac{d+2}{d}$, at least for even $d$. With the upper bound on squashed entanglement we not only match the best known upper bounds on distillable entanglement, but obtain the new bound, \eqref{eq:KD-upper}, on the distillable key, since by Lemma~\ref{lemma:squashedupper} $K_D(\alpha_d) \leq E_{\rm sq}(\alpha_d)$.

Our bound gives $E_{\rm sq}(\alpha_d) \leq \frac{2 \log_2 e}{d-1} = O(\frac{1}{d})$ which improves over the bound $E_{\rm sq}(\alpha_d) =O(\frac{\log_2 d}{d})$, obtained using the monogamy of squashed entanglement~\citep{Aaronson-squashed}. On the other hand, both $E_D$ and $K_D$ are $\geq \frac{1}{d}$. Up to a constant, the bound that we have obtained for squashed entanglement, distillable key (and distillable entanglement, but this we knew before) is therefore optimal. Previously the best upper bound for distillable key was $\half$, from a computation of the relative entropy of entanglement with respect to separable states of two copies of $\alpha_d$~\cite{vedralplenio,horodecki-2005-94,VollbrechtWerner01}.

\medskip\noindent
{\bf Lower bound on the entanglement cost.}
\label{sec:cost}
The calculation of the entanglement cost using the formula~(\ref{eq:E_F-regularized}) seems
very daunting in general due to the infinite limit; but in fact, even the computation of entanglement of formation
is a very difficult task in general. However, for 
$\alpha_d$ the $g\otimes g$ symmetry (for unitary $g$) comes to help.
\begin{lemma}
  \label{lemma:renyi-2}
  \hfill
  $E_F(\alpha_d^{\ox n}) \geq - \log_2 \displaystyle{\max_{\ket{\psi}_{A^nB^n} \in \anti^{\otimes n}}}
   \tr \psi_{A^n}^2$,
  where $\psi_{A^n}=\tr_{B^n} \proj{\psi}_{A^nB^n}$. We use Young diagrams to
  denote the subspaces of the irreducible representations of $U(d)$: 
  $\anti=\wedge^2(\complex^d)$ with associated projector $P_{\anti}$.
\end{lemma}
\begin{proof}
By definition, $E_F(\alpha_d^{\otimes n}) 
= \min_{\{p_i, \ket{\psi_i} \}: \alpha_d^{\otimes n}=\sum_i p_i \proj{\psi_i} } \sum_i p_i H(\psi_{A, i})$.
Note that all states appearing in the ensembles are contained in $\anti^{\otimes n}$. 
Thus $E_F(\rho^{\otimes n}) \geq \min_{\ket{\psi}_{A^nB^n} \in \anti^{\otimes n}} H(\psi_{A^n})$ 
(in fact this is an equality: just take any minimizer and twirl it). 
The proof follows by noting that the von Neumann entropy is lower bounded by the 
quantum R\'enyi entropy of order two, $H_2(\sigma)=-\log_2 \tr \sigma^2$.
\end{proof}

\begin{figure}
\includegraphics[width=\columnwidth]{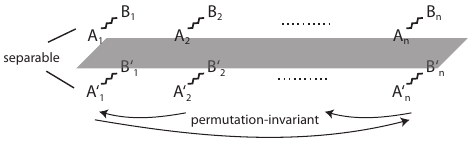}
\caption{Illustration of the properties of the state $\Omega_{A^nB^nA'^nB'^n}$. Systems connected by a curly line are in the state $\alpha_d$. Each group $A_iB_iA_i'B_i'$ is $g^{\otimes 4}$-invariant.}
\label{fig:sym}
\end{figure}

\medskip
Yura~\cite{Yura:E_C} proved that the r.h.s.~equals $n$ if $d=3$. Together with the observation that the $E_C(\alpha_d)\leq E_F(\alpha_d) = 1$,
he has thus calculated entanglement cost of $\alpha_3$. 
In the following we will show that the r.h.s.~is lower bounded by $n$ 
times $\log_2\frac{4}{3} \gtrsim 0.415$ for all $d$, and how to recover
Yura's result for $d=3$.
\begin{lemma}\label{lemma:sep}

  \(
    \displaystyle{\max_{\ket{\psi}_{A^nB^n} \in \anti^{\otimes n}}} \! \tr \psi_{A^n}^2
             \! = \! \max \tr \Omega_{A^nA'^n} F_{A^n:A'^n},
  \)
  where the second maximisation is over all states
  \begin{equation}
    \label{eq:ppt}
    \Omega_{A^nB^nA'^nB'^n} = \!\!\! \sum_{y^n \in \{ \col, \boxx, \hook\}^n} \!\!\!
                                 p_{y_1\ldots y_n}\rho_{y_1}\otimes \cdots \otimes \rho_{y_n}
  \end{equation}
 that are separable across the the $A^nB^n : A'^nB'^n$ cut and invariant under permutation of the systems $A_iB_iA'_iB'_i$. 
 The states $\rho_{y_i}$ on $A_iB_iA'_iB'_i$ are proportional to the projectors onto the spaces in \eqref{eq:decomp}
 below (see Fig.~\ref{fig:sym}).
\end{lemma}

\begin{proof}
Note that $\tr \psi_{A^n}^2 = \tr (\psi_{A^n}\otimes \psi_{A'^n}) F_{A^n:A'^n}$,
where $F_{C:D}$ is the operator that permutes (``flips'') systems $C$ and $D$. 
Since $A^n=A_1 \cdots A_n$ and likewise for $A'^n$, we have 
$F_{A^n:A'^n}=F_{A:A'}^{\otimes n}$ and therefore 
\[
  \tr \psi_{A^n}^2 = \tr (\psi_{A^nB^n}\otimes \psi_{A'^nB'^n}) (F_{A:A'}^{\otimes n} \otimes \1_{B^nB'^n}).
\]
Because $F_{A:A'}$ commutes with $g^{\otimes 2}$ for all $g \in U(d)$, we can replace 
$\psi_{A^nB^n}\otimes \psi_{A'^nB'^n} $ by the twirled state $\Omega_{A^nB^nA'^nB'^n} = \cT_{ABA'B'}^{\otimes n} (\psi_{A^nB^n}\otimes \psi_{A'^nB'^n})$ where $\cT_{ABA'B'}$ is defined by 
$\cT_{ABA'B'}(X)=\int {\rm d}g\: g^{\otimes 4}X(g^\dagger)^{\otimes 4}$ with 
the normalised Haar measure ${\rm d}g$ on $U(d)$. 
Computing the plethysms $ \sym^2(\anti) $ and $ \wedge^2(\anti) $ we find
(see App.~B~\cite{ChristandlHighly2009})
\begin{equation}\label{eq:decomp}
  \anti^{\otimes 2} \cong \sym^2(\anti) \oplus \wedge^2(\anti) 
                    \cong \left( \col \oplus \boxx\right) \oplus \hook.
\end{equation}
By elementary representation theory we can extend this result to the $n$-fold products and obtain \eqref{eq:ppt}. 
Note further that $\Omega_{A^nB^nA'^nB'^n}$ is separable across
$A^nB^n : A'^nB'^n$ and can be taken to be permutation-invariant. 
\end{proof}

\medskip
The requirement of separability is difficult to handle, so we will relax
it to the state having positive partial transpose (PPT).
At this point we are dealing with a semidefinite programme, but using
representation theory we can express the PPT condition as a linear constraint and the target function as a linear function in the variables $p_{y^n}$ ($\vec{p}:=(p_{y^n})_{y^n}$).

\begin{lemma} \label{lemma:LP}
$\displaystyle{\max_{\ket{\psi}_{A^nB^n} \in \anti^{\otimes n}}}  \tr \psi_{A^n}^2  \leq  \zeta_{n, d}$, where
\begin{equation}  
  \zeta_{n, d} := \max\, \vec{t}^{\ox n} \!\cdot \vec{p} \; \; \text{ s.t. } \vec{p} \geq 0,\  \vec{1}\cdot\vec{p} = 1,\ {T}_d^{\ox n}\vec{p} \geq 0. 
\label{eq:d-LP}
\end{equation}
Here, $\vec{t} = (-1, \half, 0)$, and the matrix $T_d$ is given by
\[
 {T}_d = 
 \left(\begin{array}{ccc}
                                  1 &      1 &             -1 \\
                  -2- \frac{6}{d-2} &      1 &  \frac{2}{d-2} \\
               1+\frac{2(d^2-d+1)}{d(d-1)(d-2)}
                                    & 1-\frac{d+1}{d(d-1)}
                                             & 1-\frac{2d-3}{d(d-1)(d-2)}
         \end{array}\right).
\]
\end{lemma}

\begin{proof} 
The objective function takes the form
\[
  \tr \Omega_{A^nA'^n} F_{A^n:A'^n} = \!\!\!\!\!\!
    \sum_{y^n \in \{ \col, \boxx, \hook\}^n} \!\!\!\!\!\!
                p_{y_1\ldots y_n} \prod_{i=1}^n t_{y_i},
\]
where we define $\tilde\rho_{y}=\tr_{BB'}\rho_y$, and
the coefficients $t_y = \tr \tilde\rho_{y} F_{A:A'}$ are the result of a straightforward
but lengthy calculation (see App.~A~\cite{ChristandlHighly2009}).
We then compute the partial transposes of $\rho_y$ with respect to $AB:A'B'$. Since these $\rho_y^\Gamma$ commute with all $g\ox g\ox\overline{g}\ox\overline{g}$, it is natural to first find
the decomposition of the space $\wedge^2(\complex^d) \otimes \wedge^2(\complex^d)$ under this action into irreducible components. 
It turns out that the space has three multiplicity-free components, 
which can be given as the supports of positive operators
(see App.~B~\cite{ChristandlHighly2009}).
The rows of ${T}_d$ contain the components of $\rho_y^\Gamma$ in terms of these operators.
$\Omega$ is PPT is then equivalent to $ {T}_d^{\ox n}\vec{p} \geq 0$.
\end{proof}
The case $d=3$ is special because $\col$ does not appear in $\anti^{\otimes 2}$. The linear programme can be solved easily and we recover Yura's result that for all $n$, $E_F(\alpha_3^{\ox n}) = n$, and hence $E_C(\alpha_3) = 1$.
For $d\geq 4$, $\col$ is present and things are more complicated.
Because of the LOCC monotonicity of $E_F$ under
twirling, $E_F(\alpha_d^{\ox n})$ is non-increasing with $d$,
so we aim to understand this linear programme for fixed $n$ but asymptotically
large $d$. In the limit $d \rightarrow \infty$,
\[
  {T}_d \longrightarrow
  {T}_\infty = \left(\begin{array}{rrr}
                 1 & 1 & -1 \\
                -2 & 1 &  0 \\
                 1 & 1 &  1
               \end{array}\right).
\]
\begin{corollary}\label{cor:simpler-LP} 
  With $T= \left(\begin{array}{rr}
                -2 & 1 \\
                1 & 1 
           \end{array}\right)$ and $\vec{t}=(-1,\half)$,
  \begin{equation}\begin{split}
      \zeta_{n, d} \leq 
      \zeta_n :=  \max\, \vec{t}^{\ox n} \!\cdot \vec{p} 
               =  2^{-n} \!\!\!\!\! \sum_{y^n\in\{\col, \boxx \}^n} \!\!\! p_{y^n}(-2)^{|y^n|}
       \label{eq:simpler-LP}
  \end{split}\end{equation}
  s.t.~$\vec{p} \geq 0$, $\vec{1} \cdot \vec{p} \leq 1$, $-T^{\ox n}\vec{p} \leq 0$,
  and $p_{y^n}$ only depends on the number $|y^n|$ of occurrences of $\col$.
\end{corollary}
\begin{proof}
Consider the linear programme~(\ref{eq:d-LP}) for $d\rightarrow\infty$ and write the constraints in the form 
$ (\vec{w}_{i_1} \otimes \cdots \otimes \vec{w}_{i_n} ) \cdot \vec{p} \geq 0$, where $\vec{w}_j$ denotes the $j$'th row of $T_\infty$. We will now drop all constraints which contain the first row of $T_\infty$, i.e.~we delete this row from $T_\infty$.
Then we see that no $y^n$ ever need to occur that contains one or more $\hook$'s. Namely, in the expansion of the state $\Omega$ every single
occurrence of $\rho_\hook$ may be replaced with 
$\frac{1}{3}\rho_\col + \frac{2}{3}\rho_\boxx$,
turning a feasible point into a new feasible point, and not
changing the value of the objective function.

It follows that we can delete the last column of $T_\infty$, as its entries never appear again in the constraints.
\end{proof}

\medskip
Now all that is left is to find an upper bound on $\zeta_n$.
\begin{lemma} \label{lemma:threequarters}
$\zeta_n\leq (\frac{3}{4})^n$, hence $E_F(\alpha_d^{\ox n}) \geq n \log_2\frac{4}{3}$.
\end{lemma}
\begin{proof}
The dual linear programme to~\eqref{eq:simpler-LP} is given by 
\begin{equation}
  \min\, z  \quad \text{ s.t. }  \vec{q} \geq 0,\   z\vec{1} - S^{\ox n}\vec{q} \geq \vec{t}^{\ox n},
  \label{eq:simpler-dual-LP}
\end{equation}
where $ S = T^\top $. 
Its value equals $\zeta_n$ by linear programming duality. 
A short calculation shows that the constraints are equivalent to the set of inequalities
\begin{equation}
  z \geq (-2)^m\, 2^{-n} + \sum_{k=0}^n \delta_k 
                                          \sum_{\ell} (-2)^\ell {m \choose \ell} {n-m \choose k-\ell}
  \label{eq:dual-symmetrised}
\end{equation}
for $m=0, \ldots, n$. It is easily checked that $z=(3/4)^n$, together with
$\delta_k = 2^{k-2n}$ for $k<n$, and $\delta_n = 0$, is dual feasible,
thus providing an upper bound on the primal.

This is the last step in the argument proving the lower bound 
on entanglement cost, \eqref{eq:EC-lower}.
\end{proof}

\medskip\noindent
{\bf Conclusion.}
\label{sec:conclusion}
We have proven a constant lower bound on the entanglement cost of the $d\times d$ antisymmetric state by way of calculating its R\'{e}nyi-2 entropic version as a convex optimisation problem, and using a linear programming relaxation. Tighter relaxations are possible, in principle 
capable of obtaining the exact value of the maximum purity of the
reduced state over all $\ket{\psi} \in \anti^{\ox n}$\footnote{In addition to the PPT condition $AB: A'B'$, one could, for instance, impose that $\Omega$ is shareable to more parties.}. 
At the same time, we showed that the squashed 
entanglement of these states, and thus the distillable key, is arbitrarily small.
We believe that our result is the strongest indication to date that
``quantum bound key'' exists: states with positive key cost to create them
(a notion not yet defined in the literature, and a little tricky
to formalize cleanly), but with zero distillable key. 

In comparison to the large gap observed between the entanglement of formation and distillable key~\cite{generic-entanglement}, our work exhibits three advantages. Firstly, our example is constructive, secondly, we show that the distillable key can be made arbitrarily small and thirdly, we 
consider the entanglement cost, which is the right measure to compare with the distillable key, and which can be strictly smaller than the entanglement of formation~\cite{Hastings}. The distinction between entanglement cost and entanglement of formation is crucial here, as it was for the discovery of bound entanglement~\cite{boundentanglement}, since the asymptotic measure of distillable key has to be compared to an asymptotic measure of preparing the state.

Finally, we can also lower bound the regularised relative entropy of 
entanglement of $\alpha_d$ w.r.t.~separable states~\cite{vedralplenio}:
\(
  E_{R,\mathrm{sep}}^\infty(\alpha_d) 
                = \! \lim_{n\rightarrow\infty} \frac{1}{n}E_{R,\mathrm{sep}}(\alpha_d^{\ox n}).
\)
Namely, by the same argument as in~\cite{audenaert-2001-87},
$E_{R,\mathrm{sep}}(\alpha_d^{\ox n}) = -\log_2 \max \tr\sigma P_{\anti}^{\ox n}$,
where the maximum is over states $\sigma$ separable across the cut $A^n:B^n$.
However,
\[\begin{split}
 & \max_{{\sigma \text{ separable} \atop \text{across }A^n:B^n}}   \tr\sigma P_{\anti}^{\ox n}
       = \max_{\ket{\alpha}\in A^n,\, \ket{\beta}\in B^n} 
                    \bra{\alpha}\bra{\beta} P_{\anti}^{\ox n} \ket{\alpha}\ket{\beta}    \\
       &\; \;  = \max_{\ket{\alpha}\in A^n,\, \ket{\beta}\in B^n \atop \ket{\psi}\in\anti^{\ox n}}
                    \bigl| \bra{\alpha}\bra{\beta} \psi \rangle \bigr|^2   = \max_{\ket{\psi}\in\anti^{\ox n}}
                    \bigl\| \tr_{B^n} \proj{\psi} \bigr\|_\infty,
\end{split}\]
and the last line is evidently upper bounded by the square root of the
maximum purity, which we showed above to be $\leq (3/4)^n$.
Hence, $E^\infty_{R,\mathrm{sep}}(\alpha_d) \geq \log_2 \sqrt{\frac{4}{3}} \gtrsim 0.207$.
In contrast, the calculation of~\cite{audenaert-2001-87} shows
$E_{R,\mathrm{PPT}}^\infty(\alpha_d) = \log_2\frac{d+2}{d}$ for the
relative entropy measure w.r.t.~PPT states. We conclude that 
squashed entanglement can be much smaller than the separable
relative entropy measure; the opposite separation was known before thanks
to the ``flower states'' of~\cite{Horodecki2005}.

\medskip\noindent
{\bf Acknowledgments.}
After completion of this work, F.~Brand\~{a}o kindly pointed out to us that 
the states from~\cite{Horodecki2005} can be used to construct states with 
$E_C(\rho)\geq \half$ and $K_D(\rho)\leq \frac{2}{\log_2 d}$. MC acknowledges the DFG (CH 843/1-1 and CH 843/2-1) and the SNF. 
NS is supported by the EU (QUEVADIS, SCALA) and the DFG (MAP).
AW is supported by the EC, the
U.K.~EPSRC, the Royal Society, and the Leverhulme Trust;
CQT is funded by the Singapore MoE and NRF.

\end{document}